\documentclass[twoside,twocolumn,aps,floatfix,prstab,eqsecnum,%
superscriptaddress]{revtex4}
\usepackage{amsfonts,amsmath,bm}
\usepackage{color}
\usepackage[dvips]{graphicx}

\begin{document}


\title{Critical Casimir forces for ${\cal O}(n)$ systems with long-range
interaction\\
in the spherical limit}

\author{H. Chamati}
\email{chamati@issp.bas.bg} \affiliation{Institute of Solid State
Physics - BAS, 72 Tzarigradsko Chauss\'ee, 1784 Sofia, Bulgaria}
\affiliation{The Abdus Salam International Centre for Theoretical
Physics, Strada Costiera 11, I-34100 Trieste, Italy}
\author{D.M. Dantchev}
\email{daniel@imbm.bas.bg}
\affiliation{Institute of Mechanics -
BAS, Academic Georgy Bonchev St. bl. 4, 1113 Sofia, Bulgaria}
\affiliation{Max-Planck Institute f\"{u}r Metallforschung,
Stuttgart, Germany}
\affiliation{Institut f\"{u}r Theoretische und
Angewandte Physik, Universit\"{a}t Stuttgart, Pfaffenwaldring 57,
D-70569 Stuttgart, Germany}
\date{\today}

\begin{abstract}
We present exact results on the behavior of the thermodynamic
Casimir force and the excess free energy
in the framework of the $d$-dimensional spherical model with a
power law long-range interaction decaying at large distances $r$
as $r^{-d-\sigma}$, where  $\sigma<d<2\sigma$ and $0<\sigma\leq2$.
For a film geometry and under periodic boundary conditions we
consider the behavior of these quantities near the bulk critical
temperature $T_c$, as well as for $T>T_c$ and $T<T_c$. The
universal finite-size scaling function governing the behavior of
the force in the critical region is derived and its asymptotics
are investigated. While in the critical and under critical region
the force is of the order of $L^{-d}$, for $T>T_c$ it decays as
$L^{-d-\sigma}$, where $L$ is the thickness of the film. We
consider both the case of a finite system that has no phase
transition of its own, when $d-1<\sigma$, as well as the case with
$d-1>\sigma$, when one observes a dimensional crossover from $d$
to a $d-1$ dimensional critical behavior. The behavior of the
force along the phase coexistence line for a magnetic field $H=0$
and $T<T_c$ is also derived. We have proven analytically that the
excess free energy is always negative and monotonically increasing
function of $T$ and $H$.
For the Casimir force we have demonstrated that for any $\sigma
\ge 1$ it is everywhere negative, i.e. an attraction between the
surfaces bounding the system is to be observed. At $T=T_c$ the
force is an increasing function of $T$ for $\sigma>1$ and a
decreasing one for $\sigma<1$. For any $d$ and $\sigma$ the
minimum of the force at $T=T_c$ is always achieved at some $H\ne
0$.
\end{abstract}
\maketitle

\section{Introduction}
\label{Intr} When a fluid is confined in a film geometry with a
thickness $L$, the boundary conditions which the order parameter
has to fulfill at the surfaces bounding the system lead to a $L$
dependence of the excess free energy.  On its turn, the last lead
to a force, conjugated  to $L$, which is called the Casimir
(solvation) force and the corresponding  effect - the
thermodynamic Casimir effect. In this form it has been discussed
for the first time by M. E. Fisher and de Gennes in 1978
\cite{Fisher78}. The effect is dubbed so after the Dutch physicist
Hendrik Casimir who first, in 1948 \cite{C48}, predicted it
considering the influence of the zero-point quantum mechanical
vacuum fluctuations of the electromagnetic field on the resulting
force between two infinite perfectly conducting planes placed
against each other. In that form the effect is known as the
quantum mechanical Casimir effect. For a long time the effect was
considered as a theoretical curiosity but the interest in it has
blossomed in the last decade. Numerous calculations and
experiments have been performed both on the thermodynamic and the
quantum Casimir effect. For a review on the thermodynamic effect
the interested  reader might consult \cite{Krech94,Kr99,BDT00},
and for the quantum one \cite{Most97,KG99,BMM01,M01}.

The Casimir force in statistical-mechanical systems  at a
temperature $T$ and in the presence of an external magnetic field
$H$ is characterized by the excess free energy due to the {\it
finite-size contributions} to the total free energy of the system.
In the case of a film geometry $L\times \infty ^2$, and under
given boundary conditions $\tau $ imposed across the direction
$L$,  the Casimir force is defined as
\begin{equation}\label{casimirdef}
F_{{\rm Casimir}}^\tau (T,H,L)=-\frac{\partial f_\tau ^{{\rm
ex}}(T,H,L)}{
\partial L},
\end{equation}
where $f_\tau ^{{\rm ex}}(T,H,L)$ is the excess free energy
\begin{equation}
f_\tau ^{{\rm ex}}(T,H,L)=f_\tau (T,H,L)-Lf_{{\rm bulk}}(T,H).
\label{eq2}
\end{equation}
Here $f_\tau (T,H,L)$ is the full free energy per unit area and
per $k_BT$, and $f_{{\rm bulk}}(T,H)$ is the corresponding bulk
free energy density. According to the standard finite-size scaling
theory \cite{BDT00,P90}, under periodic boundary conditions
$\tau=p$ near the critical point $T=T_c, H=0$ (of the bulk system)
one expects
\begin{equation}\label{fesc}
f_p ^{{\rm ex}}(T,H,L)=L^{-(d-1)}X_{f}^{(p)}(a t L^{1/\nu}, b h
L^{\Delta/\nu}),
\end{equation}
wherefrom one has
\begin{equation}\label{Csc}
F_{{\rm Casimir}}^{(p)} (T,H,L)=L^{-d}X_{{\rm Casimir}}^{(p)}(a t
L^{1/\nu}, b h L^{\Delta/\nu}).
\end{equation}
Here the universal scaling functions of the free energy
$X_f^{(p)}(x_1,x_2)$ and the Casimir force $X_{\rm
Casimr}^{(p)}(x_1,x_2)$ are related via the relation
\begin{eqnarray}\label{con}
X_{\rm Casimir}^{(p)}(x_1,x_2)&=&(d-1)X_f^{(p)}(x_1,x_2)\nonumber \\
&&-\frac{1}{\nu}x_1\frac{\partial}{\partial
x_1}X_f^{(p)}(x_1,x_2)\nonumber\\
& &-\frac{\Delta}{\nu}x_2\frac{\partial}{\partial
x_2}X_f^{(p)}(x_1,x_2),
\end{eqnarray}
$\Delta$ and $\nu$ are the standard critical exponents, $a$ and
$b$ are nonuniversal metric factors, $t=(T-T_c)/T_c$ is the
reduced temperature and $h=\beta H$ ($\beta=(k_BT)^{-1}$).
We recall that, according to
the general theory of the thermodynamic Casimir effect
\cite{Krech94,Kr99,BDT00}, $X_{\rm Casimir}^{(p)}(x_1,x_2)$ is
supposed to be negative under periodic boundary conditions (which
corresponds to a mutual attraction of the `surfaces' bounding the
system). The boundaries influence the system to a depth given by
the bulk correlation length $\xi _\infty (T)\sim |T-T_c|^{-\nu }$,
where $\nu $ is its critical exponent. When $\xi _\infty (T)\ll L$
the Casimir force, as a {\it fluctuation induced force} between
the plates, is negligible. The force becomes long-ranged when $\xi
_\infty (T)$ diverges near {\it and} below the bulk critical point
$T_c$ in an ${\cal O}(n)$, $n\geq 2$ model system in the absence
of an external magnetic field~\cite{Krech92,Danchev96,Danchev98}.
Therefore in statistical-mechanical systems one can turn on and
off the Casimir effect merely by changing, e.g., the temperature
of the system.

The temperature dependence of the Casimir force for
two-dimensional systems has been investigated exactly only on the
example of Ising strips~\cite{Evans94}. In ${\cal O}(n)$ models
for $T>T_c$ the temperature dependence of the force has been
considered in~\cite{Krech92}. The only example where it is
investigated exactly as a function of both the temperature and of
the magnetic field scaling variables is that of the
three-dimensional spherical model with short range interaction
under periodic boundary
conditions~\cite{Danchev96,Danchev98,Danchev2004}. There results
for the Casimir force in a mean-spherical model with
$L\times\infty^{d-1}$ geometry, $2<d<4$, have been derived. The
force is consistent with an {\it attraction} of the plates
confining the system. In \cite{CDT2000} some of the results of
\cite{Danchev96,Danchev98} have been extended to a quantum version
of the model. There the interaction has been taken to be
long-ranged, with $0<\sigma \le 2$, where $\sigma/2<d<3 \sigma
/2$, and the corresponding quantum phase transition has been
considered around $T=0$. Very recently in \cite{Danchev2004},
based on a derived there stress-tensor-like operator for critical
lattice systems,  the scaling functions of the force for the $3d$
Ising, XY and Heisenberg models have been obtained by Monte Carlo
methods.
The results suggest that, under periodic boundary conditions, the
scaling function $X_{\rm Casimir}^{(p)}(x)/n$ of all the $O(n)$
models practically coincide for large $x$, say, for $x=L/\xi
\gtrsim 2$, where $\xi$ is the true bulk correlation length. The
last increases the helpfulness of the spherical model results
(i.e. of the results in the limit $n\rightarrow\infty$), which are
available in an explicit analytic form.

Most of the results for the Casimir force are available only at
$T=T_c$, i.e.  for the Casimir amplitudes. They are obtained for
$d=2 $ by using conformal-invariance methods for a large class of
models~\cite{Krech94}. For $d\neq 2$ results for the amplitudes
are available via field-theoretical renormalization group theory
in $4-\varepsilon$ dimensions~\cite{Krech94,Krech92,Eisen95},
Migdal-Kadanoff real-space renormalization group
methods~\cite{Indeku}, and, by Monte Carlo
methods~\cite{Danchev2004,krech96}. In addition to the flat
geometries some results about the Casimir amplitudes between
spherical particles in a critical fluid have been derived
too~\cite{Eisen95,HSED98}. For the purposes of experimental
verification that type of geometry seems especially suitable. For
$d=3$ the only exactly known amplitude is that one for the
spherical model \cite{Danchev98}. In the case $d=\sigma$ the
amplitude is also known \cite{CDT2000} for the quantum version of
the model with long-ranged power-law interaction (in that case the
amplitude in question characterizes the leading temperature
correction to the ground state of the quantum system).

It should be noted that in contrast to the quantum mechanical
Casimir effect, that has been tested experimentally with high
accuracy \cite{L97,HCM00,MR98,BSOR02} (for a recent review on the
existing experiments see, e.g. \cite{L02}), the
statistical-mechanical Casimir effect lacks so far a
quantitatively satisfactory experimental verification.
Nevertheless, one has to stress that all the existing experiments
\cite{ML99,GC99,GC02,UBMCR03} are in a qualitative agreement with
the theoretical predictions.

In this paper a theory of the scaling properties of the Casimir
force of a spherical model with a power-law leading long-ranged
interactions (decreasing at long distances $r$ as $1/r^{d+\sigma
}$, with $0<\sigma\le 2$, and $\sigma<d<2\sigma$) is presented.
The results represent an extension to leading long-ranged
interactions of the corresponding ones for system with
short-ranged interaction \cite{Danchev96,Danchev98}. The latter
results, as we will see, can be reobtained by formally taking the
limit $\sigma \rightarrow 2^-$ in the expressions pertinent to the
case of long-ranged interactions.

All the interactions enter the exact expressions for the free
energy only through their Fourier transform which leading
asymptotic behavior is $U(q)\sim a_\sigma q^{\sigma^*}$
\cite{BDT00,joyce72}, where $\sigma^*=\min(2,\sigma)$. As it was
shown for bulk systems by renormalization group arguments
$\sigma\ge 2$ corresponds to the case of subleading long-ranged
interactions, i.e. the universality class then does not depend on
$\sigma$ \cite{FMN72} and coincides with that one of systems with
short-ranged interactions. Values satisfying $0<\sigma<2$
correspond to leading long-ranged interactions and the critical
behavior depends then on $\sigma$ (see Ref. \cite{C2001,CT2003}
and references therein). In the current work we will restrict
ourselves to the consideration of this case only. The other case
of subleading long-ranged interaction, i.e. when $\sigma>2$ is
also of interest (involving, e.g., a serious modification of the
standard finite-size scaling theory, see e.g.
\cite{CT2003,DR2000,D2001,CD2002,DKD2003}), but will be considered
elsewhere \cite{GDD}.

The investigation of the Casimir effect in a classical system with
long-range interaction possesses some peculiarities in comparison
with the short-range system. Due to the long-range character of
the interaction there exists a natural attraction between the
surfaces bounding the system. One easily can estimate that in
systems with real boundaries (i.e. with no translational
invariance) in the ordered state the $L$-dependent part of the
excess free energy that is raised by the direct inter-particle
(spin) interaction is of the order of $L^{-\sigma+1}$. In the
critical region one still has some effects stemming from that
interaction on the background of which develops the fluctuating
induced new attraction between the surfaces that  is in fact the
critical Casimir force. In the definition (\ref{casimirdef}) used
here, that is the common one when one considers short-range
systems, these effects are superposed simultaneously.  In the
current article we will investigate their interplay. An
interesting case when forces of similar origin are acting
simultaneously is that one of the wetting when the wetting layer
is nearly critical and intrudes between two noncritical phases if
one takes into account the effect of long-range correlations and
that one of the long-range Van der Waals forces
\cite{NI85,note1,KD92b}.

The structure of the article is as follows. In Section
\ref{smodel} we briefly describe the spherical model (which, in
systems with a translational invariance, is equivalent to the
$n\rightarrow\infty$ limit of the $O(n)$ models) and give all
basic expressions needed to investigate the behavior of the
Casimir force. In Section \ref{sec3} we derive the scaling
function of the excess free energy and the Casimir force, and
investigate the leading asymptotic behavior of the force both
above and below the critical point. In Section
\ref{phasecoexistence} we consider in some details the behavior of
the force along the phase coexistence line $T<T_c$, $H=0$.
In Section \ref{monotonicity} we investigate the
monotonicity properties of the excess free energy,
and the Casimir force, and prove analytically that both the excess
free energy and the force are negative for any $T$ and $H$ (for
$\sigma>1$). The last implies that the force between the boundary
surfaces of the system is always attractive. The article closes
with a discussion given in Section \ref{concl}. The technical
details needed in the main text are organized in a series of
Appendices.

\section{The model}
\label{smodel}

We consider the ferromagnetic mean spherical model with long-range
interaction confined to a fully finite $d$-dimensional hypercubic
lattice ${\cal L}_d$ of $N=|{\cal L}_d|$ \, sites. The model is
defined by
\begin{equation}\label{model}
{\cal H}=-\frac12\sum_{ij}{\cal J}_{ij}\ {\cal S}_i{\cal
S}_j-H\sum_i{\cal S}_i,
\end{equation}
where ${\cal S}_i$ is the spin variable at site $i$, ${\cal
J}_{ij}$ is the interaction matrix between spins at sites $i$ and
$j$, and $H$ is an ordering external magnetic field. The long-wave
length asymptotic form of the Fourier transform ${\cal J}({\bf
q})$ of the interaction potential ${\cal J}_{ij}$ is
$$
{\cal J}({\bf q})\approx{\cal J}({\bf 0})\left[1-\rho_\sigma
\omega_\sigma ({\bf q})\right], \ \ \ |{\bf q}|\to 0, \ \
\rho_\sigma>0.
$$
We suppose that the interaction in the system is long-ranged with
$0<\sigma<2$, i.e. $\omega_\sigma({\bf q})\simeq |{\bf
q}|^\sigma$. This corresponds to the inverse power-law behavior
${\cal J}({\bf r})\sim r^{-d-\sigma}$, for large spin separations
$r=|\bf r|$. The spins in the model under consideration obey the
spherical constraint
\begin{equation}\label{constraint}
\sum_i\langle{\cal S}_i^2\rangle=N,
\end{equation}
where $\langle\cdots\rangle$ denotes standard thermodynamic
averages taken with the Hamiltonian ${\cal H}$ and $N$ is the
total number of spins located at sites $i$ of finite hypercubic
lattice ${\cal L}_d$ of size $L_1\times L_2\times\cdots L_d=N$
(here $L_i$ are the linear sizes of the system measured in units
of the lattice constants).

Under periodic boundary conditions imposed along the finite directions
of the system, the free energy density of the model is given by
\cite{BDT00}
\begin{subequations}\label{finite}
\begin{eqnarray}\label{freeenergy}
\beta{\cal F}_{d,\sigma}\left(\beta,H,{\bf L}|{\bf \Lambda
}\right) &=& \frac12\sup_{\phi>0} \left\{ U_{d,\sigma}(\phi,{\bf L}|{\bf
\Lambda})\right.\nonumber\\
&&+\ln\left[\frac{\beta{\cal
J}(0)\rho_\sigma}{2\pi}\right]-\frac{\beta H^2}{{\cal
J}(0)\rho_\sigma\phi}\nonumber\\
&&-\beta{\cal
J}(0)\rho_\sigma\left(\phi+\frac{1}{\rho_\sigma}\right)\left.\right \},
\end{eqnarray}
where
\begin{equation}\label{Udef}
U_{d,\sigma}(\phi,{\bf L}|{\bf \Lambda})=\frac1N \sum_{\bf
q}\ln [\phi+\omega_\sigma({\bf q})].
\end{equation}
Here the vector ${\bf q}$ has the components $\{q_1,q_2,
\cdots,q_d\}$ where $q_j=2\pi n_j/L_j$ and
$n_j\in\left\{-M_j,\cdots,M_j-1\right\}$ with $M_j=L_j
\Lambda_j/(2\pi)\gg 1$ being integer numbers, and $\Lambda_j$
the cutoff at the boundaries of the first Brillouin zone along the $j$
direction.
The spherical field $\phi$ is introduced to ensure the fulfillment
of the constraint~(\ref{constraint}). It is the solution of the
equation
\begin{equation}\label{sphericalfield}
\beta{\cal J}(0)\rho_\sigma\left(1-\frac{H^2}{\phi^2{\cal J}^2(0)
\rho_\sigma^2}\right)
=\frac1N\sum_{\bf q}\frac1{\phi+\omega_\sigma({\bf q})}.
\end{equation}
\end{subequations}

Equations (\ref{freeenergy}) and (\ref{sphericalfield}) contain all the
necessary information for the investigation of the critical behavior
of the model under consideration.

In the bulk limit, when all the sizes of the system are infinite,
the $d$-dimensional sums over the momentum vector $\bf q$ in Eqs.
(\ref{Udef}) and (\ref{sphericalfield}) transform into integrals
over the first Brillouin zone. For example one has
\begin{eqnarray}\label{Ubulk}
U_{d,\sigma}(\phi| \Lambda)&=&\frac{1}{(2\pi)^d}
\int_{-\Lambda}^{\Lambda}dq_1 \cdots
\int_{-\Lambda}^{\Lambda} dq_d \nonumber\\
&&\ln [\phi+\omega_\sigma(q_1,q_2,\cdots,q_d)].
\end{eqnarray}

By analyzing the equation for the spherical field (\ref{sphericalfield})
in the bulk limit it is easy to show that the system exhibits a phase
transition for $d>\sigma$ at the critical point, $\beta_c$, given by
\begin{equation}\label{criticalpoint}
\beta_c {\cal J}(0)\rho_\sigma=
\frac{1}{(2\pi)^d}
\int_{-\Lambda}^{\Lambda}dq_1 \cdots
\int_{-\Lambda}^{\Lambda} dq_d\frac1{\omega_\sigma(q_1,q_2,\cdots,q_d)}.
\end{equation}

\section{Scaling form of the excess free energy and the critical
Casimir force}\label{sec3} In the remainder we consider a system
with a film geometry $L\times\infty^{d-1}$, which results after
taking the limits $L_2\to\infty,\cdots,L_d\to\infty$ and setting
$L_1=L$. For the simplicity of notations we will only consider the
case when all cut-off variables are taken to be equal to each
other, i.e. $\Lambda_i=\Lambda$, $i=1,\cdots,d$. Then
$U_{d,\sigma}(\phi,{\bf L}|{\bf \Lambda})$ becomes
\begin{eqnarray}\label{Ufilm}
U_{d,\sigma}(\phi, L| \Lambda)&=&\frac1L \sum_{
q_1}\frac{1}{(2\pi)^{d-1}}\int_{-\Lambda}^{\Lambda}dq_2 \cdots
\int_{-\Lambda}^{\Lambda} dq_d \nonumber\\
&&\ln [\phi+\omega_\sigma(q_1,q_2,\cdots,q_d)].
\end{eqnarray}
The above sum can be evaluated using the Poisson
summation formula and the identity
\begin{equation}\label{id}
\ln(1+z^a)=a\int_{0}^{\infty}\frac{dx}x\left(1-e^{-zx}\right)E_{a}(-x^a),
\end{equation}
where $E_a(x)\equiv E_{a,1}(x)$, and
\begin{equation}\label{mittagt}
E_{\alpha,\beta}(z)=\sum_{k=0}^{\infty}\frac{z^k}{\Gamma(\alpha
k+\beta)}
\end{equation}
are the Mittag-Leffler functions. For a review on the
properties of $E_{\alpha,\beta}(z)$ and other related to them
functions, as well as for their application in statistical and
continuum mechanics see Ref.~\cite{mainardi97} (see  also
Ref.~\cite{brankov89}). The properties used in the current article
are summarized in Appendix~\ref{appA}.

 After some
algebra for the full free energy density we receive:
\begin{equation}\label{fss}
\beta {\cal F}_{d,\sigma}(\beta,H,L)=\beta{\cal F}_{d,\sigma}(\beta,H)-
\frac12L^{-d}{\cal K}_{d,\sigma}(L^\sigma\phi),
\end{equation}
where
$$
{\cal F}_{d,\sigma}(\beta,H)\equiv
\lim_{L\to\infty}{\cal F}_{d,\sigma}(\beta,H,L),
$$
and
\begin{eqnarray}\label{kfun}
{\cal K}_{d,\sigma}(y)&=&\frac{\sigma}{(4\pi)^{d/2}}\sum_{l=1}^{\infty}
\int_0^\infty dx x^{-d/2-1}\exp\left(-\frac{l^2}{4x}\right)\nonumber\\
&&\times E_{\sigma/2,1}\left(-x^{\sigma/2}y\right),
\end{eqnarray}
The main advantage of the above expression
for the free energy, despite its complicated form in comparison to
equation (\ref{freeenergy}), is the simplified dependence on the
size $L$ which now enters only {\it via} the arguments of some
functions. This gives the possibility, as it is explained below,
to obtain the scaling functions of the excess free energy and the
Casimir force. It is worthwhile noting that under a sharp cutoff
$\Lambda$ a special care has to be taken when performing
finite-size scaling calculations in order to avoid receiving
artificial, i.e. not existing in real systems, finite-size
$\Lambda$-dependent contributions. This question is considered in
details in \cite{DKD2003}. In obtaining (\ref{fss}) the suggested
there receipt has been applied (see Eq. (27) in \cite{DKD2003} and
the discussion devoted to it). According to these findings, for
the finite-size contributions in the following we are going to
send the cutoff to infinity.

In equation (\ref{fss}), $\phi$ is the solution of the
corresponding spherical field equation that follows by requiring
the partial derivative of the right hand side of equation
(\ref{fss}) with respect to $\phi$ to be zero.
Let us denote the
solution of the corresponding bulk spherical equation by
$\phi_\infty$. Then, for the excess free energy (per unit area) it
is possible to obtain from equations (\ref{eq2}) and (\ref{fss})
the finite size scaling form, valid for $\sigma<d<2\sigma$,
\begin{equation}\label{excess}
f^{\text ex}(\beta,H,L|d)
=\beta L^{-(d-1)}X_f\left(x_1,x_2\right),
\end{equation}
with scaling variables
\begin{subequations}\label{sv}
\begin{equation}
x_1=\left(\beta-\beta_c\right){\cal J}(0)\rho_\sigma L^{1/\nu}
\end{equation}
and
\begin{equation}
x_2=HL^{\Delta/\nu}\sqrt{\beta/{\cal J}(0)\rho_\sigma}.
\end{equation}
\end{subequations}
Here $\nu=1/(d-\sigma)$ and $\Delta=(d+\sigma)/(2(d-\sigma))$ are
the critical exponents of the spherical model (for
$\sigma<d<2\sigma$, and $0<\sigma\le 2$).
In equation (\ref{excess}) the universal scaling function $X^{\text
ex}\left(x_1,x_2\right)$ of the excess free energy has the form
\begin{eqnarray}\label{excessscaling}
&&X_f\left(x_1,x_2\right)=-\frac12x_2^{ 2}\left(
\frac1{y_L}-\frac1{y_\infty}\right)
-\frac12x_1\left(y_L-y_\infty\right)\nonumber\\
&&-\frac\sigma{2d}\left|D_{d,\sigma}\right|\left(y_L^{d/\sigma}
-y_\infty^{d/\sigma}\right) -\frac{1}{2}{\cal K}_{d,\sigma}(y_L),
\end{eqnarray}
where the $y_L=\phi_L L^\sigma$,
$y_\infty=\phi_\infty L^\sigma$, and
\begin{equation}\label{constD}
D_{d,\sigma}=2\pi\left[(4\pi)^{d/2}\Gamma\left(\frac d2\right)
\sigma\sin\left(\frac{\pi d}\sigma\right)\right]^{-1}.
\end{equation}

In Eq. (\ref{excessscaling}) $y_L$ is the solution of the
spherical field equation for the finite system obtained by
minimizing the free energy with respect to $y_L$
\begin{equation} \label{fses}
x_1=\frac{x_2^2}{y_L^2}-|D_{d,\sigma}|y_L^{d/\sigma-1}
-\frac\partial{\partial y_L}{\cal K}_{d,\sigma}(y_L).
\end{equation}
For the infinite system the corresponding equation is
\begin{equation}\label{ises}
x_1=\frac{x_2^2}{y_\infty^2}-|D_{d,\sigma}|
y_\infty^{d/\sigma-1}.
\end{equation}

According to equation (\ref{casimirdef}), the finite-size scaling
function of the Casimir force for the system under consideration
is
\begin{eqnarray}\label{Casimir}
&&X_{\text{Casimir}}\left(x_1,x_2\right)=\frac{\sigma+1}{2}x_2^{2}\left(
\frac1{y_L}-\frac1{y_\infty}\right)\nonumber\\
&&-\frac{\sigma-1}{2}x_1\left(y_L-y_\infty\right)
-\frac{\sigma(d-1)}{2d} |D_{d,\sigma}| \left(y_L^{d/\sigma}-
y_\infty^{d/\sigma}\right)\nonumber\\
&&-\frac12(d-1){\cal K}_{d,\sigma}(y_L).
\end{eqnarray}

Note that in the limit $\sigma\to 2^-$ Eqs.
(\ref{excess})-(\ref{Casimir}) reproduce exactly the corresponding
ones for the case of the short-range interaction
\cite{Danchev96,Danchev98,Danchev2004}. In such a case the above
equations simplify greatly since then $E_{1,1}(z)=\exp(z)$, and
the function ${\cal K}_{d,\sigma}(y)$ defined in Eq. (\ref{kfun})
becomes
\begin{equation}
{\cal K}_{d,2}(y)=\frac{4}{(2\pi)^{d/2}}y^{d/4}\sum_{l=1}^\infty
l^{-d/2}K_{d/2}(l\sqrt{y}),
\end{equation}
where $K_{\nu}$ is the modified Bessel function.

In the present article we will concentrate on the investigation of
the behavior of the Casimir force and the excess free energy  in
different regions of the phase diagram. We will also evaluate some
critical amplitudes for selected values of the parameters $d$ and
$\sigma$. The analysis will be done analytically for the cases
where one can obtain simple expressions and is then extended
numerically to cases which are not accessible by analytical means.

\begin{figure}[ht]
\resizebox{\columnwidth}{!}{\includegraphics{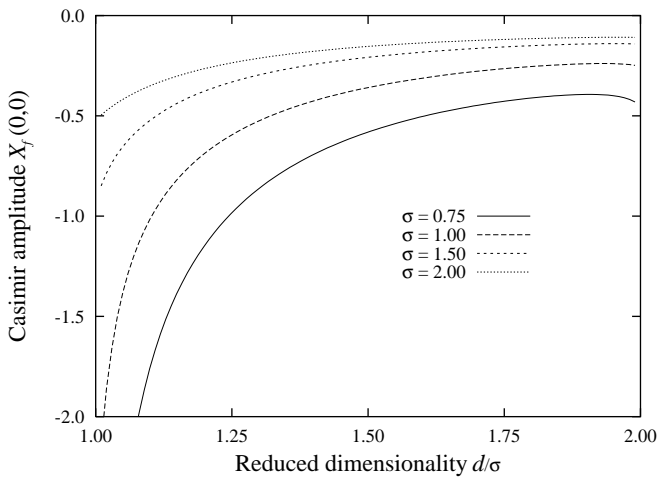}}
\caption{Behavior of the Casimir amplitude as a function of  $d$.}
\label{casampl}
\end{figure}

In FIG. \ref{casampl} we present the numerical evaluation of the
Casimir amplitudes as a function of $d$ for some selected values
of $\sigma$. The results show that the amplitude is an
increasing function of $d$ at fixed $\sigma$, and an increasing
function of $\sigma$ at a fixed $d$. Note also that in accordance
with the general expectations, the amplitudes are negative.

\begin{figure}[ht]
\resizebox{\columnwidth}{!}{\includegraphics{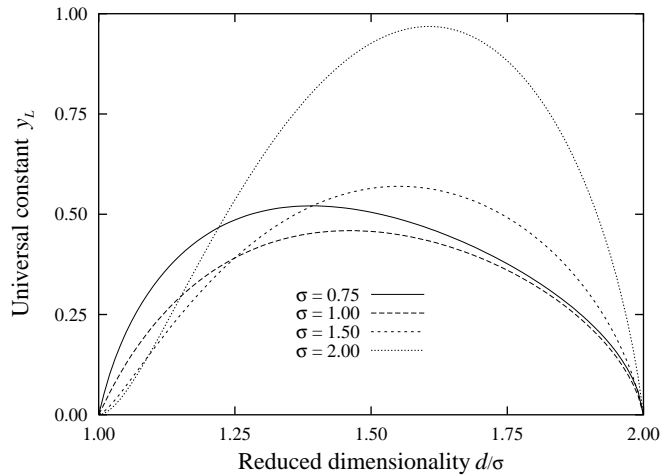}}
\caption{Behavior of the scaling variable $y_L$ as a function of
$d$ at the critical point $T=T_c$.  We recall that the finite-size
correlation length $\xi_L$ is related to $y_L$ via
$\xi_L=Ly_L^{-1/\sigma}$ \cite{BDT00}. } \label{corlength}
\end{figure}

In order to obtain the amplitudes, one needs to know the value of
$y_L(T)$ at the critical point $T=T_c$ that is the solution of the
equation for the spherical field (\ref{fses}). These results have
their own important physical meaning. We recall that $y_L$ is
directly connected to the finite-size correlation length
$\xi_L=Ly_L^{-1/\sigma}$ of the system \cite{BDT00}. The results
for $y_L(T_c)$ are shown in FIG. \ref{corlength}.

\begin{figure}[ht]
\resizebox{\columnwidth}{!}{\includegraphics{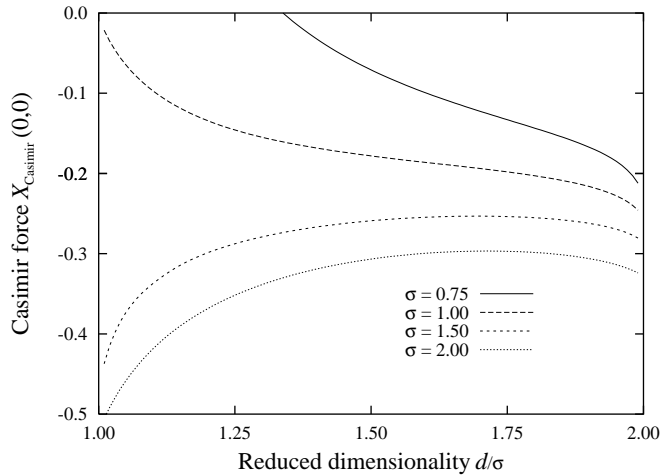}}
\caption{The behavior of the Casimir force at $T=T_c$ as a
function of $d$.}\label{forced}
\end{figure}

In FIG. \ref{forced} we present our results for the Casimir force
evaluated at the bulk critical point of the model as a function of
$d$ for some selected values of $\sigma$. We observe that the
Casimir force behaves in a different way depending on whether
$\sigma$ is smaller or larger than $\sigma=1$. For $\sigma\leq 1$
it is decreasing monotonically as a function of $d$, while for
$\sigma>1$ it is not.

In the following we turn our attention to the investigation of the
thermodynamic functions of interest as a function of the scaling
variable $x_1$ for fixed $d$ and $\sigma$. Let us first consider
the situations where it is possible to obtain some results
analytically.

Let us first consider the asymptotic of the excess free energy and
the Casimir force in the under critical region (i.e. $T \lesssim
T_c$). Taking into account that then (i.e. when $x_1 \gg 1$,
$x_2=0$), according to Eqs. (\ref{fses}) and (\ref{ises}) $y_L
\rightarrow 0^+$, $y_\infty=0$, as well as the asymptotic
(\ref{asK}) of the function ${\cal K}_{d,\sigma}(y_L)$ for small
values of the argument (derived in Appendix \ref{asyK}) it is easy
to see that below the critical temperature
\begin{equation}
\label{asXfbelow}
 X_f\left(x_1\rightarrow \infty,0\right)\simeq
-\frac{\sigma}{2\pi^{d/2}} \Gamma\left(\frac d2\right)\zeta(d),
\end{equation}
and
\begin{equation}\label{asXCbelow}
X_{\text{Casimir}}\left(x_1\rightarrow\infty,0\right)\simeq -
\frac{\sigma (d-1)}{2\pi^{d/2}} \Gamma\left(\frac
d2\right)\zeta(d).
\end{equation}
The above results reflect the dominating contribution of the
Goldstone modes in the under critical-regime of an $O(n)$ model -
both the excess free energy and the Casimir force do not  tend
exponentially-in-$L$ to zero, but to  {\it finite} constants. For
$\sigma=2$ these constants coincide with those known from
short-range systems (see, e.g. \cite{Danchev2004} and the
references cited therein). Note also that, in contrast with
systems with real boundaries, the direct inter-spin long ranged
interaction below $T_c$ does not lead to a $L^{-(\sigma-1)}$
contribution, which is well known from studies of van der Waals
systems exhibiting wetting phase transitions \cite{NI85,Di88}.
This is due to the application of periodic boundary conditions,
i.e. the system under consideration lacks real physical
boundaries.


Let us consider the critical behavior of the force  for  $T>T_c$
in a bit more details. Then, when $x_2=0$ and
$x_1\rightarrow-\infty$ from (\ref{fses}) and (\ref{ises}) one
obtains $y_L \simeq y_\infty (1+\varepsilon_{d,\sigma})$, where
\begin{equation}\label{varepsilon}
\varepsilon_{d,\sigma}=\frac{a_{d,\sigma}}{\left(\frac{d}{\sigma}-
1\right)|D_{d,\sigma}|\ y_\infty^{d/\sigma+1}},
\end{equation}
and
\begin{equation}\label{yiabove}
y_\infty=\left(\frac{|x_1|}{|D_{d,\sigma}|}\right)^{\frac{\sigma}{d-\sigma}}.
\end{equation}
Therefore, the leading behavior of the scaling function of the
force in that region is
\begin{eqnarray}\label{AsCasAbove}
X_{\rm Casimir}&\simeq& -A_{d,\sigma} y_\infty^{-1}\nonumber\\
&\simeq&-A_{d,\sigma}\left[(\beta_c-\beta)
{\cal J}(0)\rho_\sigma/|D_{d,\sigma}|\right]^{-\frac{\sigma}{d-\sigma}}
L^{-\sigma},\nonumber\\
&&
\end{eqnarray}
where
\begin{equation}\label{Ads}
A_{d,\sigma}=\frac{a_{d,\sigma}}{2}\left(\sigma+d-1\right).
\end{equation}
Eq.~(\ref{AsCasAbove}) implies that above $T_c$, $F_{\rm Casimir}
\simeq -X_+ |t|^{-\gamma}L^{-(d+\sigma)}$, with
$\gamma=\sigma/(d-\sigma)$, and $X_+
>0$, i.e. the force remains attractive and decays in a
power-in-$L$ and not in an exponentially-in-$L$ way, as it is in
systems with short ranged interactions. This behavior is in full
correspondence with the long-ranged character of the interaction.
Similar is, as it has been recently established, also the behavior
of the Casimir force and the excess free energy in systems with
van der Waals type interaction \cite{DKD2003} (see also
\cite{GDD}), despite that their critical exponents are that ones
of the short-ranged systems.

\begin{figure}[ht]
\resizebox{\columnwidth}{!}{\includegraphics{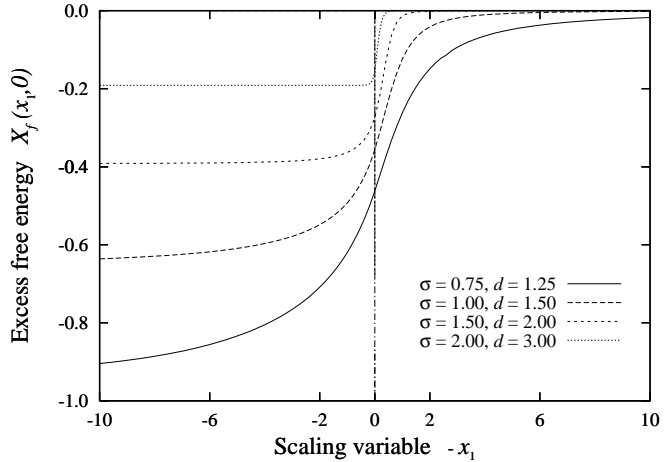}}
\caption{The universal finite-size scaling function of the excess
free energy $X_f(x_1,0)$ from Eq. (\ref{excessscaling}) as a
function of $-x_1\sim (T-T_c)L^{1/\nu}$, for some selected values
of $\sigma$ at zero external magnetic field. One observes that, in
full accordance with the corresponding statement from Section
\ref{monotonicity}, $X_{\rm Casimir}(x_1,0)$ is a monotonically
increasing function of the temperature $T$ regardless of value of
$\sigma$.} \label{excessx1}
\end{figure}

\begin{figure}[ht]
\resizebox{\columnwidth}{!}{\includegraphics{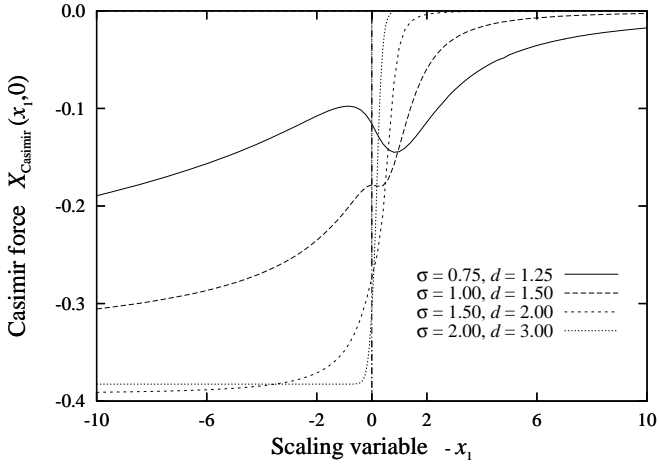}}
\caption{The universal finite-size scaling function of the Casimir
force $X_{\rm Casimir}(x_1,0)$  as a function of the scaling
variable $-x_1\sim (T-T_c)L^{1/\nu}$,   at zero external magnetic
field $H=0$. One observes that, in full accordance with the
corresponding statement from Section \ref{monotonicity}, $X_{\rm
Casimir}(x_1,0)$ is a monotonically increasing function of the
temperature $T$ (for $\sigma>1$) and possesses a complex behavior
for $\sigma\le 1$.} \label{forcex1}
\end{figure}

The obtained analytical results are supported by numerical
analysis of the expressions for the scaling functions of the
excess free energy and the Casimir force at zero external field.
The corresponding data is presented in FIG. \ref{excessx1} (for
the excess free energy) and in FIG. \ref{forcex1} (for the Casimir
force). While the scaling function of the excess free energy is
monotonic regardless of the used values of $d$ and $\sigma$, the
behavior of Casimir force depends strongly on the range of the
interaction $\sigma$. For $\sigma>1$ it is monotonically
increasing as it can be seen from the case $\sigma=2$,
corresponding to short range interaction, and the long-range case
with $\sigma=1.5$. For $\sigma=1$ the monotonicity changes and
$X_{\rm Casimir}(x_1,0)$ becomes decreasing for values of
$\sigma<1$. As example we show its behavior for $\sigma=0.75$.

\begin{figure}[ht]
\resizebox{\columnwidth}{!}{\includegraphics{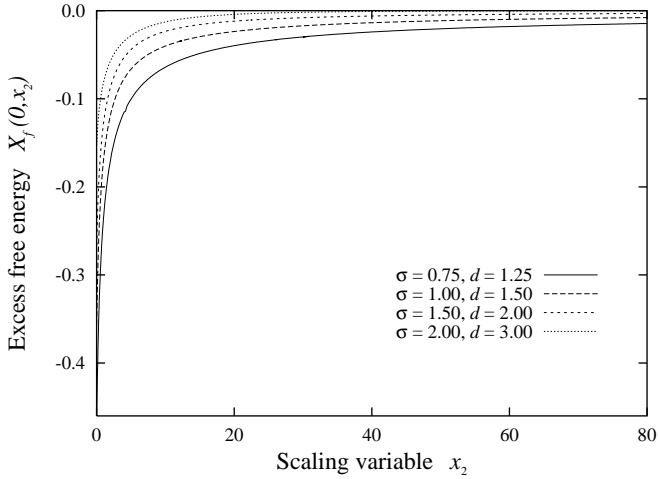}}
\caption{The universal finite-size scaling function of the excess
free energy $X_{\rm f}(0,x_2)$, for some values of $\sigma$, as a
function of the scaling variable $x_2\sim HL^{\Delta/\nu}$ at the
bulk critical point $T=T_c$. One observes $X_f(0,x_2)$ is a
monotonically increasing function of the field $H$ for arbitrary
$\sigma$.} \label{excessx2}
\end{figure}

\begin{figure}[ht]
\resizebox{\columnwidth}{!}{\includegraphics{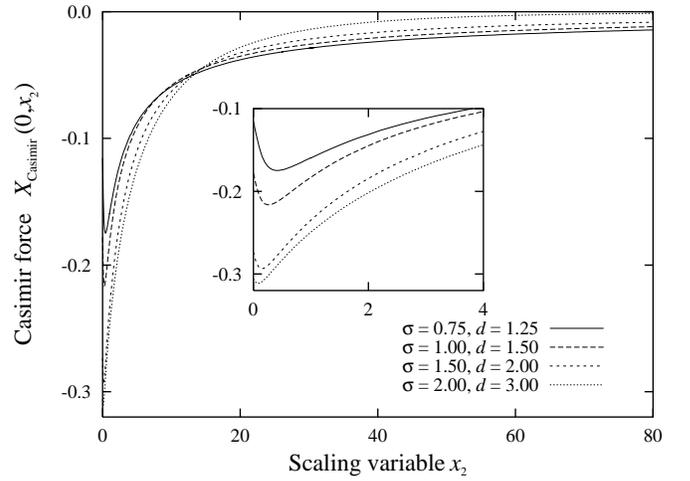}}
\caption{The universal finite-size scaling function of the Casimir
force $X_{\rm Casimir}$, for some values of $\sigma$, as a
function of the scaling variable $x_2\sim HL^{\Delta/\nu}$ at the
bulk critical temperature $T=T_c$. One observes that, $X_{\rm
Casimir}(0,x_2)$ is not a monotonically increasing function of the
field $H$ for all values of $\sigma\leq2$ including the
short-range case.} \label{forcex2}
\end{figure}

We close this section by presenting the outcome of the numerical
analysis of the behavior of the scaling functions of the excess
free energy, shown in FIG. \ref{excessx2}, and that of the Casimir
force, shown in FIG. \ref{forcex2}, as a function of the scaling
variable $x_2$ at the bulk critical temperature. One observes that
the excess free energy is a monotonically increasing function of
the external magnetic field $H$ independently of the range of the
interaction. However the Casimr force is a {\it nonmonotonic}
function of $H$ and {\it has a minimum at $x_2 \ne 0$} which depth
depends of the parameter $\sigma$. The minimum is found to be at
$x_2=$ 0.084, 0.145, 0.263 and 0.416 for $\sigma$ = 2, 1.5, 1 and
0.75, respectively. So, as long as $\sigma$ goes smaller the
minimum becomes deeper. Indeed the ratio of the Casimir force
evaluated at the minimum to its value at $H=0$ is a decreasing
function of $\sigma$. It is given by 1.017, 1.073, 1.215 and 1.513
for $\sigma$ = 2, 1.5, 1 and 0.75, respectively.

\section{Casimir (solvation) force along the phase coexistence line}
\label{phasecoexistence}

Here we investigate the behavior of the Casimir force along the
line $H=0$ when $T<T_c$. This is a line of a first order phase
transition with respect to the magnetic field $H$. The finite-size
rounding of the first-order transitions in $O(n)$ models has been
already studied by Fisher and Privman in \cite{FP85} for a fully
finite and cylinder geometries. Later their predictions have been
verified in details for the spherical model system with such a
geometry in \cite{FP86}, while in \cite{BD90,BD91} their arguments
have been extended to a geometry of the type $L^{d-d'}\times
\infty^{d'}$, where $d'$ has been chosen so that no phase
transition of its own  exists in the finite system, i.e.
$d'<\sigma$ has been supposed. Here we extend these investigations
to cover also the cases $d'=\sigma$ and $d'>\sigma$ in systems
with a film geometry, i.e. when $d'=d-1$. We will be only
interested in the behavior of the Casimir force.

For $T<T_c$ and small $H$ Eqs. (\ref{excess})-(\ref{Casimir}) are
still valid, but there the limit $y_L\ll 1$ has to be taken (i.e.
we suppose that $x_1 \gg x_2^2$). As it is clear from Eq.
(\ref{asK}), then there are three subcases  to be considered.

{\it i)} {\it The case $d-1<\sigma$.}

Then in the finite system there is no phase transition on its own.
For the excess free energy one obtains
 \begin{eqnarray}\label{fexbTc}
f^{\rm ex}(\beta, H)&=&-\frac{\sigma}{2
\pi^{d/2}}\Gamma(d/2)\zeta(d)L^{-(d-1)}\nonumber\\
&&+\beta m_0 H L\left\{1-\frac{1}{2}
\left(\frac{m_0}{m_L}+\frac{m_L}{m_0}\right)\right.\nonumber\\
&&\left.+\frac{\sigma}{2(d-1)}
\left(\frac{m_0}{m_L}-\frac{m_L}{m_0}\right)\right\},
\end{eqnarray}
where
\begin{equation}\label{ml}
\frac{m_L}{m_0}=\sqrt{\left[\frac{|D_{d-1,\sigma}|}{2x_m}\right]^2+1}
-\frac{|D_{d-1,\sigma}|}{2x_m}.
\end{equation}
Here $m_L=H/[\rho_\sigma {\cal J}(0) \phi]$ is the magnetization
of the finite system, $m_0=\sqrt{1-T/T_c}\ $  is the spontaneous
magnetization, and $x_m=\beta m_0(T) L \xi_L^{d-1}H$, which has
the meaning of the ratio of the total magnetic energy in the
correlated volume $V_{{\rm cor}}=L\xi_L^{d-1}$ to the thermal
energy $k_B T$ per degree of freedom, is the scaling variable. (We
recall that in the spherical model the true finite-size
correlation length $\xi_L$ is equal to $\phi^{-1/\sigma}$
\cite{BDT00,BD91}.) Next, it is easy to see from Eq. (\ref{ml})
that $x_m=O(1)$ involves $H=O(L^{-\sigma/(1+\sigma-d)})$, that is
the scale on which the jump in the bulk magnetization is rounded
off. From this observation and from Eq. (\ref{fexbTc}) one
obtains that the $H$ dependent correction to the Casimir force is
then of the order of $L^{-\sigma/(1+\sigma-d)}$ (Note that
$\sigma/(1+\sigma-d)>d$ for $d>\sigma$, and, so, the term
proportional to $H$ in Eq. (\ref{fexbTc}) will indeed contribute
as a correction towards the Casimir force).

{\it ii)}  {\it The case $d-1=\sigma$.}

This is the borderline case between that one when in the finite
system there is no phase transition of its  own (for $d-1<\sigma$)
and that one in which in the finite system there is such a phase
transition (for $d-1>\sigma$). In this case an essential singular
point exists in  the finite-size system at $T=H=0$. For the excess
free energy one now obtains
\begin{eqnarray}\label{fexbTcborderline}
f^{\rm ex}(\beta, H)&=&-\frac{\sigma}{2
\pi^{d/2}}\Gamma(d/2)\zeta(d)L^{-(d-1)}\nonumber\\
&&+\beta m_0 H L \left\{1-\frac{m_0}{m_L} \right\},
\end{eqnarray}
where
\begin{equation}\label{mlB}
\frac{m_L}{m_0}=\sqrt{\left[\frac{1}{(4\pi)^{\sigma/2}\Gamma(\sigma/2)}
\frac{1}{\bar{x}_m}\right]^2+1}+
\frac{1}{(4\pi)^{\sigma/2}\Gamma(\sigma/2)}\frac{1}{\bar{x}_m}
\end{equation}
and $\bar{x}_m=\beta m_0(T)HL\xi_L^{d-1}/\ln(L/\xi_L)$. The above
equations are to be compared with the previous case. One observes,
that the  main difference is the existence of logarithmic-in-$L$
dependence that is introduced via the scaling field variable
$\bar{x}_m$. As a result the rounding of the jump in the
magnetization takes place on a scale given by
$H=L^{-\sigma}\exp(-{\rm const.}\ L)$, i.e. the scale in this case
is exponentially small in $L$.

{\it ii)}  {\it The case $d-1>\sigma$.}

In this case there is a true phase transition of its own in the
finite system at some $T_{c,L}=T_c-\varepsilon L^{-1/\nu}$, i.e.
no rounding of the jump of the magnetization is possible. One only
observes $L$-dependent corrections of the finite-size
magnetization $m_L$ with respect to the spontaneous magnetization
$m_0$. One finds that the crossover from $d$ to $d-1$ critical
behavior happens at $T_{c,L}$ with
\begin{equation}
\label{shift} \varepsilon=\frac{\pi^{(d-1)/2}}{(2\pi)^\sigma}
\frac{C_{d,\sigma}}{\Gamma(d/2)}\frac{1}{\beta_c {\cal
J}(0)\rho_\sigma},
\end{equation}
and, when $|H|L^\sigma \ll 1$,
\begin{equation}\label{fexbTcsharp}
f^{\rm ex}(\beta, H)=-\frac{\sigma}{2
\pi^{d/2}}\Gamma(d/2)\zeta(d)L^{-(d-1)}+\beta m_0 H L
\frac{a}{L^{d-\sigma}},
\end{equation}
with
\begin{equation}\label{a}
a=\frac{\pi^{(d-1)/2}}{2(2\pi)^\sigma}
\frac{C_{d,\sigma}}{\Gamma(d/2)}\frac{1}{\beta m_0^2(T) {\cal
J}(0)\rho_\sigma},
\end{equation}
and $m_L\simeq m_0(1-a/2)$.

Finally, we would like to note that in $O(n)$ systems one observes
for $T<T_c$ in addition to the rounding of the jump of the order
parameter also rounding of the spin wave singularities. According
to the general theory \cite{FP85,FP86}, their rounding occurs on
the scale for which $x_s=|H|L^{\sigma}=O(1)$. As it is clear from
Eq. (\ref{sv}) (and taking into account that if $T<T_c$ one can
rewrite $x_1$ as $x_1=\beta m_0(T)^2 \rho_\sigma {\cal J}(0)$,
with $x_1 \gg 1$) the scale on which the rounding of the spin wave
singularities sets in involves that $x_1\sim x_2^2$ there. Then,
in this regime, the solution of the spherical field equations for
the finite and the infinite system (\ref{fses}) and (\ref{ises})
will be again $y_L=O(1)$ and $y_\infty=O(1)$. Since $x_1$ and
$x_2$ can be expressed from Eqs. (\ref{fses}) and (\ref{ises}) in
terms of $y_L$ and $y_\infty$, we conclude that, according to
(\ref{Casimir}), in the regime in which the spin waves are of
importance, the Casimir force will be $F_{\rm Casimir}=O(L^{-d})$,
possessing a nontrivial $H$ dependence. If one would like to
reveal more on this dependence the numerical treatment is
unavoidable. Note that when the field is strong enough to suppress
the spin-wave excitations, i.e. when $x_s\gg 1$ and $T<T_c$, one
will have an Ising-like system. In this regime $y_L\gg 1$,
$y_\infty\gg 1$, and the Casimir force will be of the order of
$L^{-(d+\sigma)}$ (see Eq. (\ref{AsCasAbove})) under periodic
boundary conditions. (If the system was possessing real bounding
surfaces like, say, under Dirichlet-Drichlet boundary conditions,
one would expect that the corresponding contribution in the force
is of the order of $L^{-\sigma}$.)

\section{Monotonicity properties of the excess free energy and the
Casimir force}
\label{monotonicity}


Let us denote by $g_L(x_2,y)$ and $g_\infty(x_2,y)$ the right-hand
side of Eqs. (\ref{fses}) and (\ref{ises}), respectively. Now we
prove that

{\it i)} $g_L(x_2,y) > g_\infty(x_2,y)$ and

{\it ii)} that $g_L(x_2,y)$ and $g_\infty(x_2,y)$ are
monotonically decreasing functions of $y$.

{\it i)} First, let us note that $E_{\alpha,\beta}(-x)$ is a
completely  monotonic function of $x \ge 0$
\cite{P48,MS97,S96,MS01} for $0<\alpha\le 1$ and $\beta\ge\alpha$.
(In \cite{P48}  this property was shown to hold for
$E_{\alpha,1}(-x)\equiv E_{\alpha}(-x)$ and was later extended to
$E_{\alpha,\beta}(-x)$ in \cite{MS97} and \cite{S96}; see also
\cite{MS01}.) This means that for all $n=0,1,2,3, \cdots$ one has
\begin{equation}\label{monot}
(-1)^n \frac{d^n E_{\alpha,\beta}(-x)}{dx^n}\ge 0, \ x\ge 0, \
0<\alpha \le 1, \ \beta \ge \alpha.
\end{equation}
Then, from $n=0$ it immediately follows that
$E_{\alpha,\alpha}(-x)>0$ when $x\ge 0$. Now, from Eqs.
(\ref{fses}) and (\ref{ises}), it immediately follows that
$g_L(x_2,y) > g_\infty(x_2,y)$.

{\it ii)} The required property follows from the monotonicity of
the function $E_{\alpha,\alpha}(-x)$ for $x\ge 0$ and the explicit
form of the right hand sides of Eqs. (\ref{fses}) and
(\ref{ises}).

\begin{figure*}[htb]
\begin{center}
\begin{tabular}{cc}
\resizebox{\columnwidth}{!}{\includegraphics{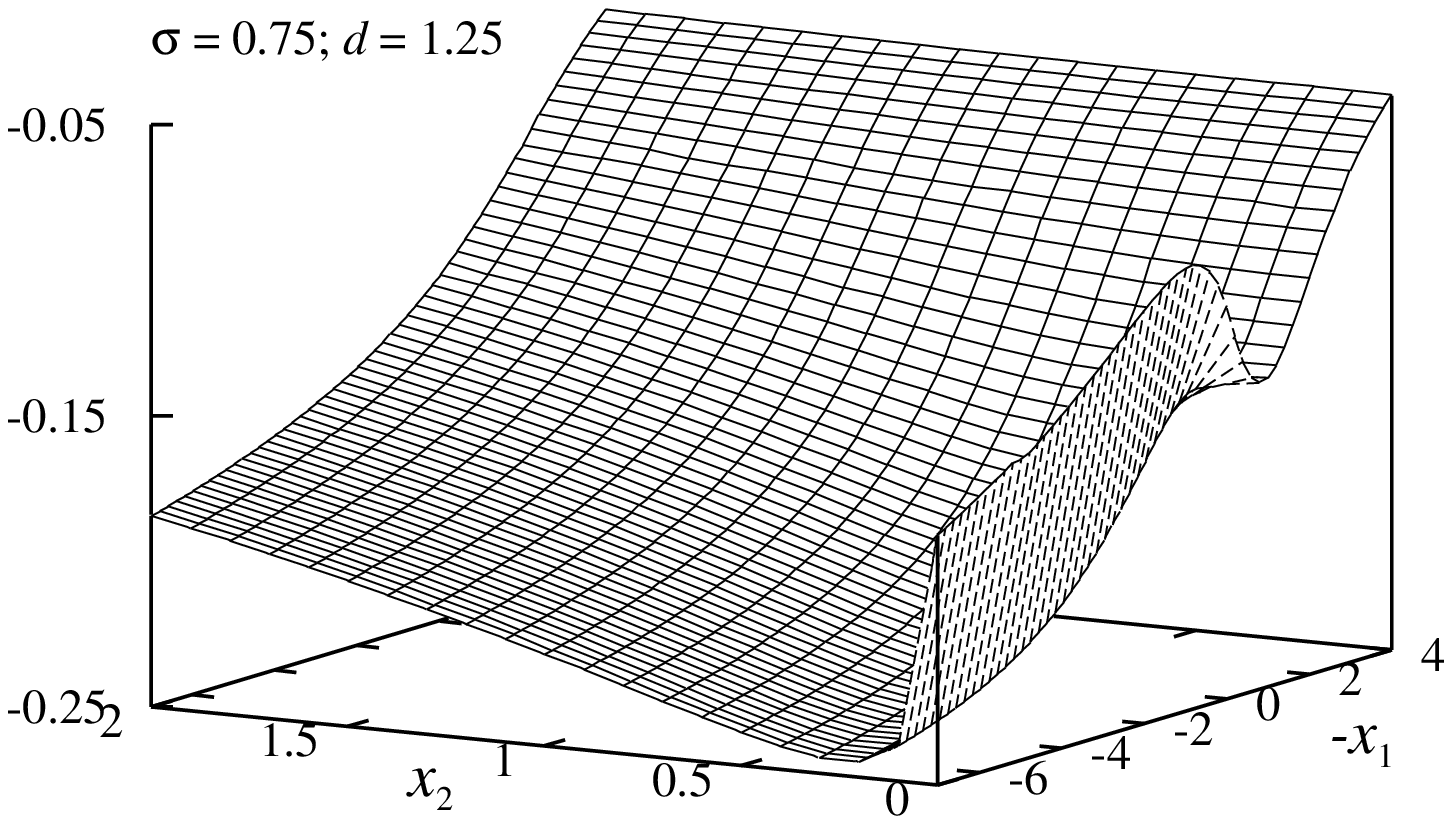}} &
\resizebox{\columnwidth}{!}{\includegraphics{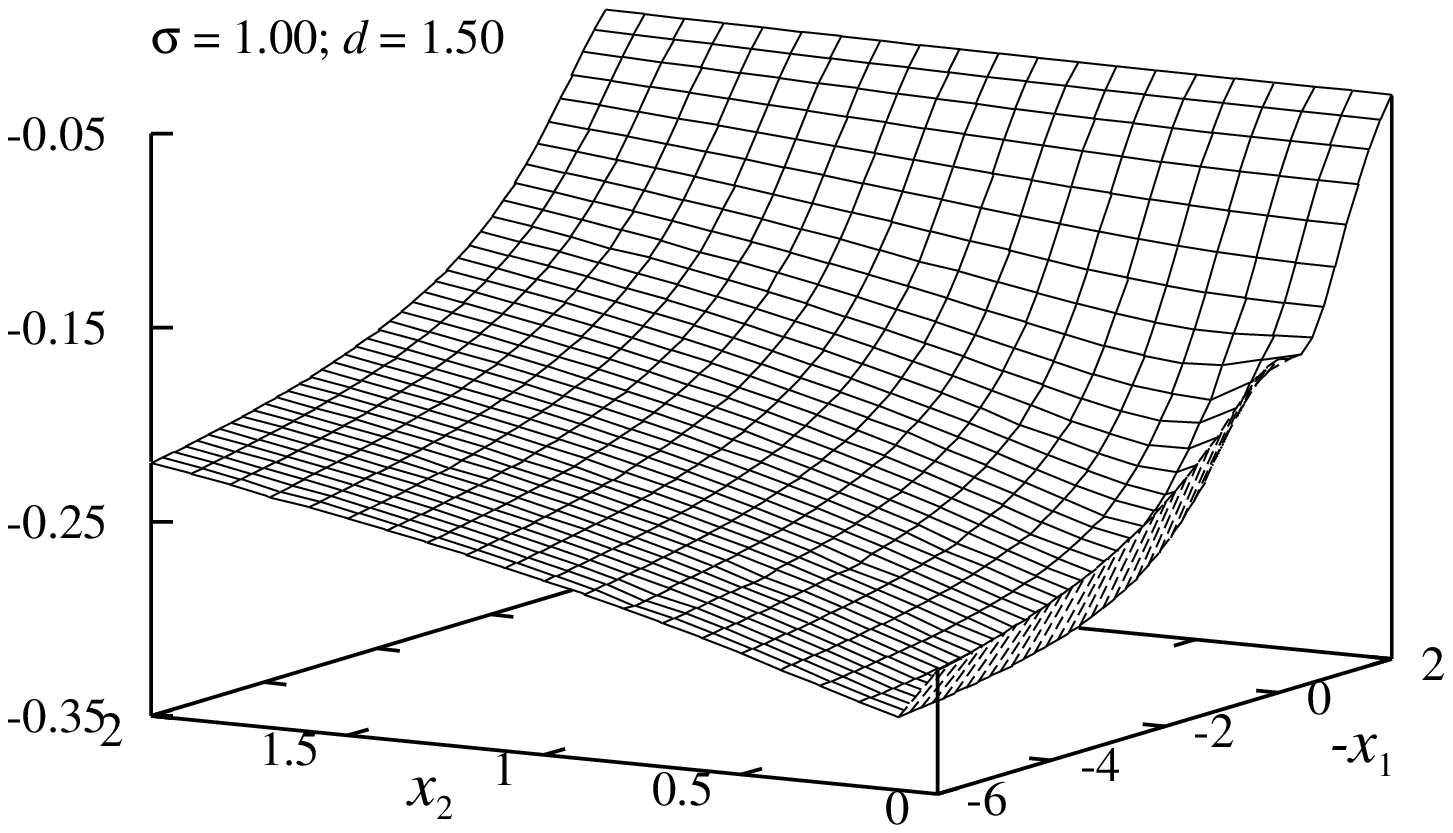}} \\[0.5cm]
\resizebox{\columnwidth}{!}{\includegraphics{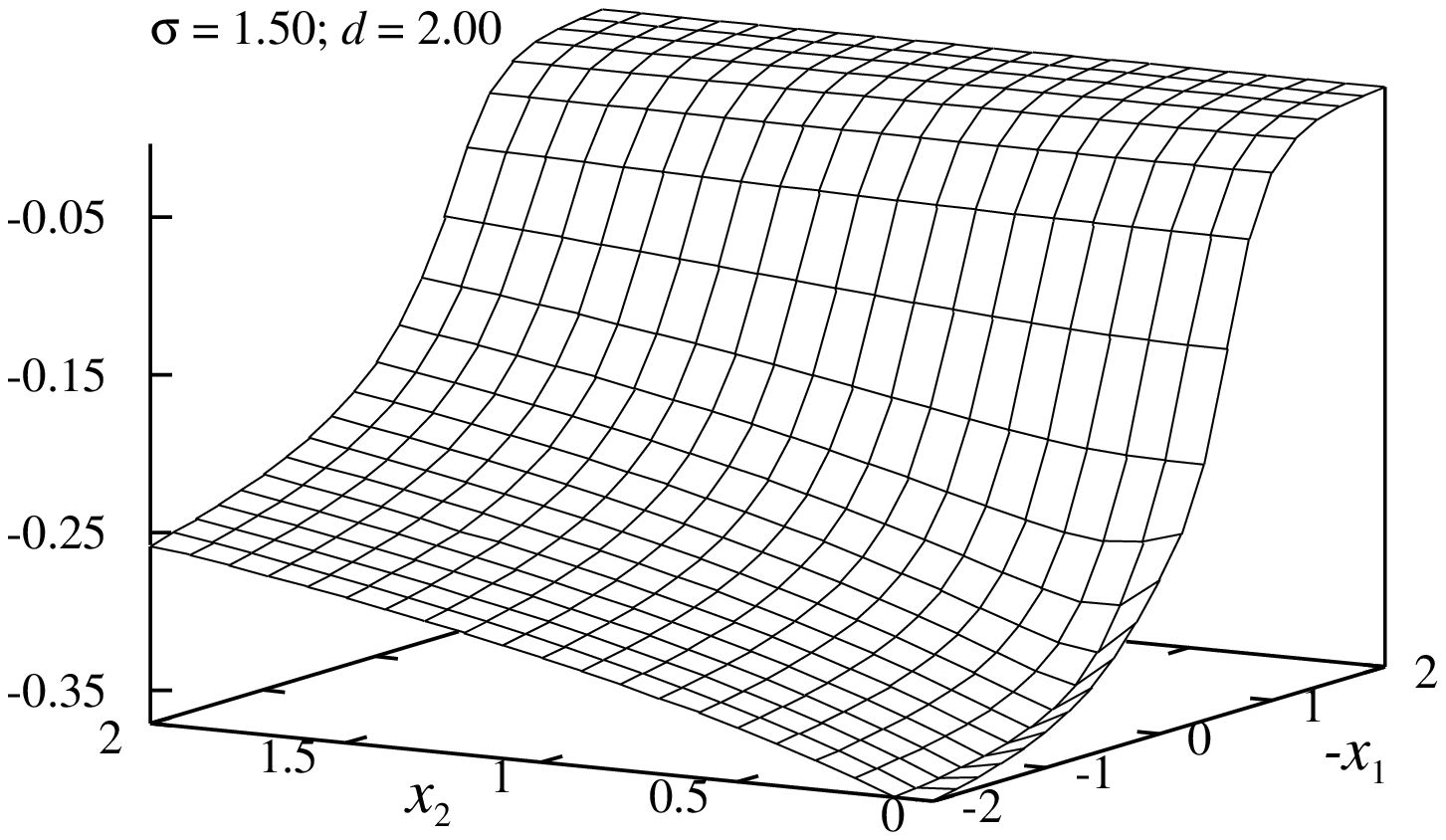}} &
\resizebox{\columnwidth}{!}{\includegraphics{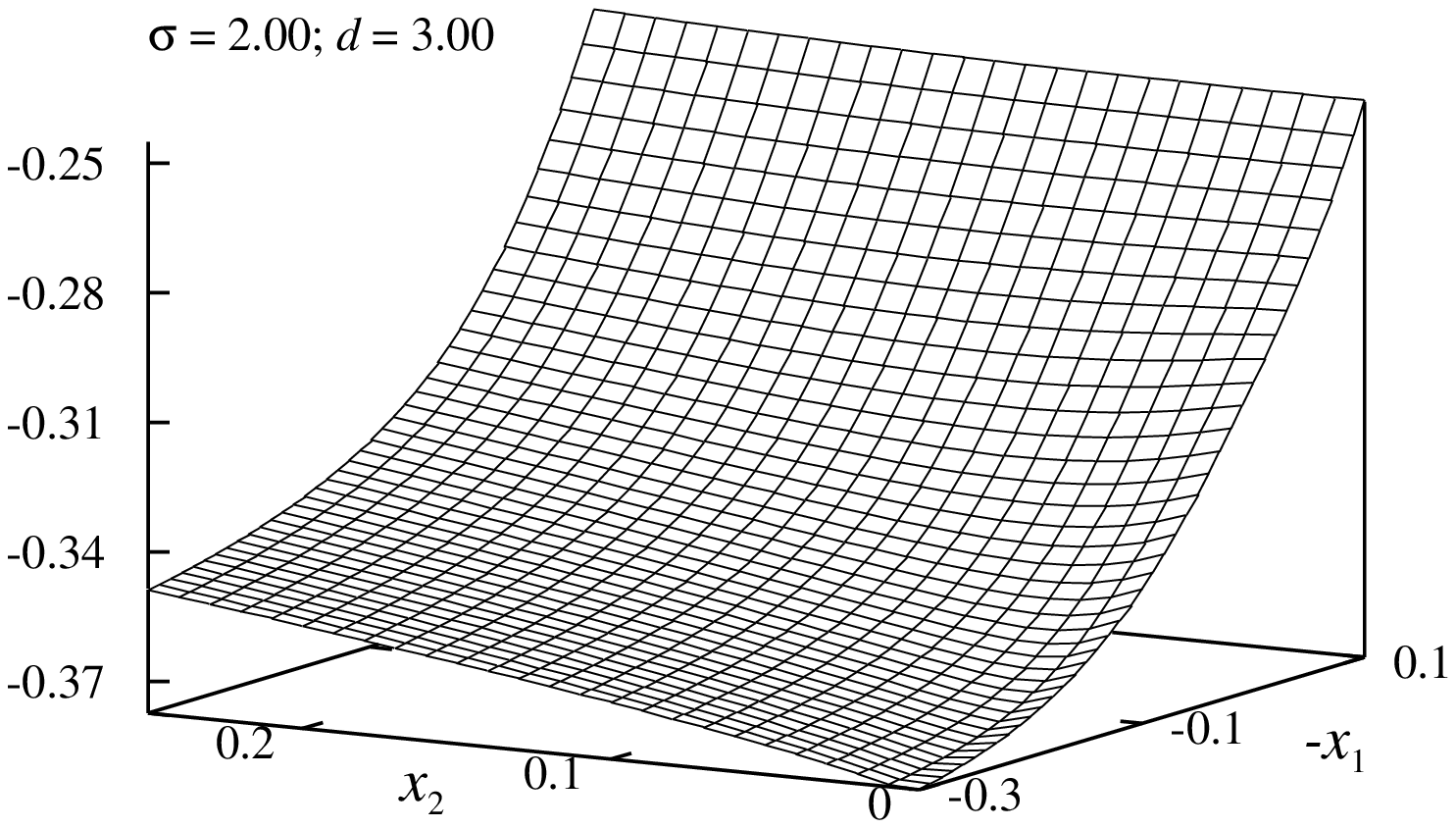}}
\end{tabular}
\end{center}
\caption{The universal finite-size scaling function of the Casimir
force as a function of scaling variables $x_1$ and $x_2$ for some
values of the parameter $\sigma$ and the corresponding values of
$d$. The visualization is limited to positive values of $x_2$
since the function is even in $H$.} \label{casimir3d}
\end{figure*}

Having proved $i)$ and $ii)$, it is easy to understand now that
for any given values $x_1$ and $x_2$ of the scaling variables the
solution of the spherical field equation for the finite system
will be larger than that for the infinite system, i.e.
$y_L(x_1,x_2)>y_\infty(x_1,x_2)$. (Since the correlation lengths
in the finite and the infinite system are $\xi_L=y_L^{-1/\sigma}$
and $\xi_\infty=y_\infty^{-1/\sigma}$ \ \cite{BDT00},
correspondingly, the physical meaning of the above result is that
the correlation length of the finite system is always smaller than
that of the infinite one.) We are then ready to prove that

{\it A) For $x_1\ge 0$ and $x_2=0$ the excess free energy scaling
function is negative, i.e. $X_{f}(x_1\ge 0,x_2=0)<0$.}

{\it B) The excess free energy scaling function $X_f(x_1,x_2)$ is
a monotonically increasing function of the temperature $T$ and the
magnetic field $|H|$.}

Let us start with statement {\it A)}.

{\it A)} From the explicit form  of the Eq. (\ref{excessscaling})
it is clear that the statement {\it A)} will be true if
$E_{\alpha,1}(-x)\ge 0$ when $x\ge0$. The last follows from
(\ref{monot}) for $n=0$, and, thus, $X_{f}(x_1,x_2)<0$.

Let us now prove  the statement {\it B}.

{\it B)} From Eq. (\ref{excessscaling}) one obtains
\begin{equation}
\frac{\partial X_f}{\partial x_1}=\frac 1 2 (y_\infty-y_L)<0,
\label{dx1}
\end{equation}
and
\begin{equation}
\frac{\partial X_f}{\partial x_2}=x_2
(\frac{1}{y_\infty}-\frac{1}{y_L})>0. \label{dx2}
\end{equation}
Eq. (\ref{dx1}) implies that {\it $X_f(x_1,x_2)$ is a monotonically
increasing function of \ $T$}, whereas Eq. (\ref{dx2}) states that
{\it it is a monotonically increasing function of $|H|$} too.

Using {\it B)} one can now prove that:

{\it C) The excess free energy scaling function is negative for
any $T$ and $H$, i.e. $X_f(x_1,x_2)<0$ for any $x_1$ and $x_2$.}

Indeed, from the monotonicity property {\it B)} and from {\it A)}
it is clear that in order to prove {\it C)} it is enough to show
that it holds for values of $T$ {\it above} $T_c$, i.e. when $y_L
\gg 1$ and $y_\infty \gg 1$. Then, from Eqs. (\ref{fses}),
(\ref{ises}) and the asymptotic (\ref{largeyasy}) one obtains
$y_L=y_\infty(1+\varepsilon)$, $0<\varepsilon \ll 1$, where
\begin{equation}\label{epsilon}
\varepsilon=\frac{a_{d,\sigma}}{y_\infty^2\left[2
\frac{x_2^2}{y_\infty^2} +\left|D_{d,\sigma}\right|
y_\infty^{d/\sigma-1}(\frac{d}{\sigma}-1)+2
\frac{a_{d,\sigma}}{y_\infty^2}\right]}.
\end{equation}
Next, from Eq. (\ref{excessscaling}) it follows that
\begin{equation}\label{exscalingas}
    X_f(x_1,x_2)\simeq -\frac{1}{2}
    \frac{a_{d,\sigma}}{y_\infty}(1-\varepsilon)<0.
\end{equation}
Thus, the excess free energy is indeed always negative.


Finally, we prove that

{\it D) For $\sigma \ge 1$ the Casimir force is always negative,
i.e. it is a force of attraction between the surfaces bounding the
system.}

We start by multiplying Eq. (\ref{ises}) with $y_\infty$ and Eq.
(\ref{fses}) with  $y_L$, and then adding the results together.
One obtains
\begin{eqnarray}\label{he}
x_1(y_L-y_\infty)&=&x_2^2(\frac{1}{y_L}-\frac{1}{y_\infty})-|D_{d,\sigma}|
\left(y_L^{d/\sigma}-y_\infty^{d/\sigma}\right)\nonumber\\
&&-y_L \frac{d}{d y_L}{\cal K}_{d,\sigma}(y_L).
\end{eqnarray}
Inserting the above expression in Eq. (\ref{Casimir}), one obtains
\begin{eqnarray}\label{CasimirF}
&&X_{\text{Casimir}}\left(x_1,x_2\right)= x_2^2\left(
\frac1{y_L}-\frac1{y_\infty}\right)\nonumber\\
&&-\frac{1}{2}\left(1-\frac{\sigma}{d} \right) |D_{d,\sigma}|
\left(y_L^{d/\sigma}- y_\infty^{d/\sigma}\right)
-\frac12(d-1){\cal K}_{d,\sigma}(y_L)\nonumber\\
&&+\frac{\sigma-1}{2}y_L
\frac{d}{d y_L} {\cal K}_{d,\sigma}(y_L).
\end{eqnarray}
Since, according to what already has been proven, $y_L>y_\infty$,
and ${\cal K}_{d,\sigma}(y_L)$ is a positive and monotonically
decreasing function of $y_L$ (the last follows from the explicit
form of ${\cal K}_{d,\sigma}(y_L)$ given in Eq. (\ref{kfun}) and
the property (\ref{monot}) of $E_{\alpha,1}(x)$ for $n=0$ and
$n=1$), from the above expression one immediately confirms the
validity of statement {\it D)}. In addition, from Eq.
(\ref{Casimir}) it is easy to see that
$X_{\text{Casimir}}\left(x_1,x_2\right)<0$ also for $\sigma<1$ if
$x_1\le 0$, i.e. for $T \ge T_c$. Furthermore, from Eqs.
(\ref{con}) and (\ref{Casimir}) it follows that
\begin{equation}\label{monCasimir}
\frac{\partial}{\partial
x_1}X_{\text{Casimir}}\left(x_1=0,x_2=0\right)=-\frac{\sigma-1}{2}y_{L,c},
\end{equation}
where from we conclude, that at $T=T_c$ the Casimir force is an
increasing function of $T$ for $\sigma>1$ (see Fig.
\ref{forcex1}), and a decreasing function of $T$ when $\sigma<1$
(see Fig. \ref{forcex1}). Therefore, at the critical point the
monotonicity of the force changes as a function of $\sigma$ at
$\sigma=1$ where we have an inflexion point.

\section{Discussion}
\label{concl}

In the current article we consider the behavior of the excess
finite-size free energy and the Casimir (solvation) force
in a classical system
with leading long range interactions in the limit
$n\rightarrow\infty$ of the $O(n)$ models (i.e. within the
spherical model). The dimensionality $d$ and the parameter
controlling the range of the interaction $\sigma$ are chosen so,
that the hyperscaling is valid, i.e. $\sigma<d<2\sigma$ is
supposed. In this regime the critical exponents depend on
$\sigma$. We demonstrate that, despite of the choice of $\sigma$,
the excess free energy scaling function $X_f$ (see FIG.
\ref{excessx1} and FIG. \ref{excessx2})
is a monotonic function of the temperature $T$ and the magnetic field
$H$, with
$X_f$ being always a negative function. Surprisingly, to a given
extend, the above properties do not hold in such a general fashion
for the Casimir (solvation) force (see FIG. \ref{forcex1} and FIG.
\ref{forcex2}). This is in line with the results of Section
\ref{monotonicity} where we show analytically that the force is
attractive for any $T$ and $\sigma \geq 1$, as well as for any $T
\geq T_c$ if $\sigma<1$. The monotonicity of the force turns out
to depend on $\sigma$. For example, if $\sigma>1$ at $T=T_c$ and
$H=0$ the force is an increasing function of $T$ and $L^{-1}$,
while for $\sigma<1$ it is a decreasing function of both $T$ and
$L^{-1}$ at this point (see Eq. (\ref{monCasimir}) and FIG.
\ref{forcex1}). In addition, one derives that for  $T=T_c$ the
minimum of the force is {\it not} at $H=0$ (see FIG.
\ref{forcex2}). Indeed, at $T=T_c$ the minimum has been found to
be at some {\it finite} value of the scaling field variable
$x_2\sim HL^{\Delta/\nu}$. For $\sigma$ = 2, 1.5, 1 and 0.75 the
minimum  at $T=T_c$ is found to be at $x_2\simeq$ 0.084, 0.145,
0.263 and 0.416, respectively. Such an occurrence of a force
minimum for a nonzero bulk field has also been reported for the
case of $(+,+)$ boundary conditions \cite{SHD03,DME00}. Here, in
this Section, we provide more details for the universal
finite-size scaling function of the Casimir force $X_{\rm
Casimir}(x_1,x_2)$ presenting the numerical results for it as a
function of both $x_1$ and $x_2\geq0$   in FIG. \ref{casimir3d}.
There the effects due to both the temperature and the magnetic
field are demonstrated (we recall that $x_1\sim (T-T_c)L^{1/\nu}$,
$x_2\sim HL^{\Delta/\nu}$). We observe that for $T<T_c$ and $H\ne
0$ a cavity shows up in the vicinity of the critical temperature
that disappears for temperatures far away from the critical point.
More precisely, one observes that there exists a finite value
$x_1^*$ of $x_1$, such that for any $x_1^*>x_1 \ge 0$ there is a
local minimum of the force at some finite $x_{2,{\rm min}}$, i.e.
at $H \ne 0$. For $x_1>x_1^*$ there is no such minimum at nonzero
$H$. In FIG. \ref{casimir3d} the last is shown for the cases
$\sigma=0.75, 1, 1.5$ and $\sigma=2$ (the short-range case). Note,
that for $\sigma=0.75$ one needs to go deeply in the under
critical region to find out where exactly the cavity vanishes. In
the short-range case $\sigma=2$ we established that $x_1^* \simeq
0.28$.

\begin{acknowledgments}
H. Chamati acknowledges financial support from the Associateship
Scheme of the Abdus Salam International Centre of Theoretical
Physics (ICTP), Trieste, Italy.

D. Dantchev acknowledges the  hospitality of Max-Planck-Institute
for Metals Research in Stuttgart.
\end{acknowledgments}

\appendix
\section{Some properties of the Mittag-leffler type functions}
\label{appA}
The Mittag-leffler type functions are defined by
the power series~\cite{mainardi97}:
\begin{equation}\label{mittag}
E_{\alpha,\beta}(z)=\sum_{k=0}^\infty\frac{z^k}{\Gamma(\alpha k+\beta)},
\ \ \ \alpha,\beta>0.
\end{equation}
They are entire functions of finite order of growth. The functions
are named after Mittag-Leffler who first considered  the
particular case $\beta=1$. These function are very popular in the
field of fractional calculus (for a recent review see
Ref.~\cite{mainardi97}).

One of the most useful property of these functions is the
identity~\cite{mainardi97}
\begin{equation}\label{ide}
\frac1{1+z}=\int_0^\infty dx e^{-x}x^{\beta-1}E_{\alpha,\beta}
\left(-x^\alpha z\right),
\end{equation}
which is obtained by means of term-by-term integration of the
series~(\ref{mittag}). The integral in Eq.~(\ref{ide}) converges in the
complex plane to the left of the line $\text{Re}\ z^{1/\alpha}=1$, $|\arg
z|\leq\frac12\alpha\pi$. The identity (\ref{ide}) lies in the basis of the
mathematical investigation of finite-size scaling in the spherical model
with algebraically decaying long-range interaction
(see Ref. \cite{BDT00} and references therein).

In some particular cases the functions $E_{\alpha,\beta}(z)$
reduce to known functions. For example, in the case corresponding
to the short range case we have
\begin{equation}\label{simple}
E_{1,1}(z)=\exp(z).
\end{equation}

Setting $z=y^{-\alpha}$, $y>0$, and $x=ty$, we obtain the Laplace
transform
\begin{equation}\label{ide1}
\frac{y^{\alpha-\beta}}{1+y^\alpha}=\int_0^\infty dt e^{-y
t}t^{\beta-1} E_{\alpha,\beta}\left(-t^\alpha\right)
\end{equation}
from which we derive the usefull identity
\begin{equation}
\frac1{1+z^\alpha}=\int_0^\infty dx \exp\left(-xz\right)x^{\alpha-1}
E_{\alpha,\alpha}\left(-x^\alpha\right),
\end{equation}
by setting $\beta=\alpha$.

The asymptotic behavior for large arguments of the Mittag-Leffler
functions is given by the Lemma~\cite{bateman55}:

Let $0<\alpha<2$, $\beta$ be an arbitrary complex number, and $\gamma$
be a real number obeying the condition
$$
\frac12\alpha\pi<\gamma<\min\{\pi,\alpha\pi\}.
$$
Then for any integer $p\geq1$ the following asymptotic expressions
hold when $|z|\to\infty$:
\begin{itemize}
\item At $|\arg z|\leq\gamma$,
\begin{equation}
E_{\alpha,\beta}(z)=\frac1\alpha z^{(1-\beta)/\alpha}e^{z^{1/\alpha}}
-\sum_{k=1}^\infty\frac{z^{-k}}{\Gamma(\beta-\alpha k)} +{\cal
O}\left(|z|^{-p-1}\right).
\end{equation}
\item At $\gamma\leq|\arg z|\leq\pi$,
\begin{equation}\label{largez}
E_{\alpha,\beta}(z)=-\sum_{k=1}^\infty\frac{z^{-k}}
{\Gamma(\beta-\alpha k)} +{\cal O}\left(|z|^{-p-1}\right).
\end{equation}
\end{itemize}

\section{Asymptotics of the function
${\cal K}_{\lowercase{d},\sigma}(\lowercase{y})$} \label{asyK}

Here we will evaluate the asymptotic behaviors of the auxiliary function
${\cal K}_{d,\sigma}(y)$ used in the expression of the free energy and
the quantities descending from it. It is defined by:
\begin{subequations}
\begin{eqnarray}\label{f}
{\cal K}_{d,\sigma}(y)&=&\frac{\sigma}{2(4\pi)^{d/2}}
\int_0^\infty dx x^{-\frac d2-1}\left[{\cal A}
\left(\frac1{4x}\right)-1\right]\nonumber\\
&&\times E_{\frac\sigma2,1}\left(-x^{\frac\sigma2}y\right),
\end{eqnarray}
where
\begin{equation}\label{A}
{\cal A}(u)=\sum_{l=-\infty}^\infty e^{-ul^2}.
\end{equation}
\end{subequations}

Using the identity
\begin{equation}
E_{\alpha,1}(-z)=1-z E_{\alpha,\alpha+1}(-z),
\end{equation}
it is possible to write down Eq. (\ref{f}) in a more convenient
form, which will allow us to extract the asymptotics of the
function under investigation. After some algebra one obtains
\begin{subequations}
\begin{equation}
{\cal K}_{d,\sigma}(y)=\sigma\pi^{-d/2}
\Gamma\left(\frac d2\right)\zeta(d)-\frac\sigma2{\cal I}_{d,\sigma}(y),
\end{equation}
where we have introduced the auxiliary function
\begin{eqnarray}\label{I}
{\cal I}_{d,\sigma}(y)&=&\frac{y}{(4\pi)^{d/2}}
\int_0^\infty dx x^{\frac\sigma2-\frac d2-1}
\left[{\cal A}\left(\frac1{4x}\right)-1\right]\nonumber\\
&& \times
E_{\frac\sigma2,\frac\sigma2+1}\left(-x^{\frac\sigma2}y\right).
\end{eqnarray}
\end{subequations}

Now, setting $x=z(2\pi)^{-2}$  and with the help of the identity
\begin{equation}
{\cal A}(u)=\sqrt{\frac\pi u}{\cal A}\left({\frac{\pi^2} u}\right),
\end{equation}
we rewrite equation (\ref{I}) (after some algebra) in the form
\begin{widetext}
\begin{eqnarray}\label{i2}
{\cal I}_{d,\sigma}(y)&=&y\frac{\pi^{(d-1)/2}}{(2\pi)^\sigma}
\int_0^\infty dx x^{\frac\sigma2-\frac d2-\frac12}
\left[{\cal A}(x)-\sqrt{\frac\pi x}-1\right]
E_{\frac\sigma2,\frac\sigma2+1}\left(-y\frac{x^{\frac\sigma2}}
{(2\pi)^\sigma}\right)\nonumber\\
&&+y\frac{\pi^{(d-1)/2}}{(2\pi)^\sigma}
\int_0^\infty dx x^{\frac\sigma2-\frac d2-\frac12}
E_{\frac\sigma2,\frac\sigma2+1}\left(-y\frac{x^{\frac\sigma2}}
{(2\pi)^\sigma}\right).
\end{eqnarray}
\end{widetext}

The integral in the second term of the right-hand side of (\ref{i2})
can be evaluated exactly with the help of the identities
\begin{equation}
u^{-\nu}=\frac1{\Gamma[\nu]}\int_0^\infty dtt^{\nu-1}e^{-ut}
\end{equation}
and
\begin{equation}
\int_0^\infty u^{\mu-1}\ln(a+b u^\nu)=\left(\frac ab\right)^{\mu/\nu}
\frac\pi{\sin(\pi\mu/\nu)}
\end{equation}
to yield the result
\begin{equation}\label{st}
2(d-1)^{-1}D_{d-1,\sigma}y^{(d-1)/\sigma}.
\end{equation}

For the evaluation of the first integral in the right hand side of
(\ref{i2}), we note that the two terms in the square brackets in
(\ref{i2}) cannot be integrated separately, since they diverge.
Nevertheless, it is possible to outwit this divergence, by
transforming further (\ref{i2}) by adding and subtracting from the
function $E_{\alpha,\alpha+1}(z)$ its asymptotic behavior at small
arguments, leading, after some algebra, to
\begin{subequations}
\begin{equation}
2\frac y\sigma\frac{\pi^{(d-1)/2}}{(2\pi)^\sigma}
\frac{C_{d,\sigma}}{\Gamma[\sigma/2]} -2d^{-1}D_{d,\sigma}y^{d/\sigma}
+{\cal R}_{d,\sigma}(y).
\end{equation}
Here we introduced the notations
\begin{equation}\label{madelung}
C_{d,\sigma}=\int_0^\infty dx x^{\frac\sigma2-\frac
d2-\frac12} \left[2\sum_{l=1}^\infty e^{-xl^2}-\sqrt{\frac\pi
x}\right], \ \ \ \ \ d-1<\sigma,
\end{equation}
and
\begin{eqnarray}
&&{\cal R}_{d,\sigma}(y)=2y\frac{\pi^{(d-1)/2}}{(2\pi)^\sigma}
\sum_{l=1}^\infty
\int_0^\infty dx x^{\frac\sigma2-\frac d2-\frac12}e^{-xl^2}\nonumber\\
&&\times\left[E_{\frac\sigma2,\frac\sigma2+1}\left(-y\frac{x^{\frac\sigma2}}
{(2\pi)^\sigma}\right)-\frac1{\Gamma[\frac\sigma2+1]}\right].
\end{eqnarray}
\end{subequations}

Collecting the above results, we obtain
\begin{eqnarray}\label{finalK}
&&{\cal K}_{d,\sigma}(y)=\sigma\pi^{-d/2}
\Gamma\left(\frac d2\right)\zeta(d)\nonumber\\
&&-\sigma(d-1)^{-1}D_{d-1,\sigma}y^{(d-1)/\sigma}
-y\frac{\pi^{(d-1)/2}}{(2\pi)^\sigma}\frac{C_{d,\sigma}}{\Gamma[\sigma/2]}
\nonumber\\
&&+\sigma d^{-1}D_{d,\sigma}y^{d/\sigma}
-\frac\sigma2{\cal R}_{d,\sigma}(y).
\end{eqnarray}

The constant $C_{d,\sigma}$ introduced in Eq. (\ref{madelung}) is the
so called Madelung constant (see e.g. \cite{chamati1996,chamati2000})
\begin{eqnarray}\label{madelung1}
C_{d,\sigma}&=&\lim_{\delta\to0}\left\{
2\sum_{l=1}^\infty\frac{\Gamma[(\sigma-d+1)/2,\delta l^2]}
{l^{(\sigma-d+1)/2}}\right.\nonumber\\
&&\left.-\int_{-\infty}^\infty dl\frac{\Gamma[(\sigma-d+1)/2,\delta l^2]}
{l^{(\sigma-d+1)/2}}\right\}, d-1<\sigma,\nonumber\\
&&
\end{eqnarray}
where $\Gamma[\alpha,x]$ is the incomplete gamma function. It has
been shown that this constant has a remarkable property of
symmetry \cite{chamati2000}, which relates its values in the case
$d-1<\sigma$ to those in the case $d-1>\sigma$. On the other hand,
it has been shown that $C_{d,\sigma}$ can be expressed in terms of
the analytic continuation, over $d-1<\sigma$, of (for details see
\cite{chamati2000})
\begin{equation}\label{zeta}
C_{d,\sigma}=2\pi^{\frac12+\sigma-d}
\Gamma\left(\frac{d-\sigma}2\right)\zeta(d-\sigma), \ \ \ \ \
d-1>\sigma.
\end{equation}

Eq. (\ref{finalK}) is the general form of the functions ${\cal
K}_{d,\sigma}(y)$. According to Eqs. (\ref{madelung1}) and
(\ref{zeta}) it can be used to investigate the critical behavior
of the system for any dimension less than $d$.

For small $y$ the asymptotic behavior of the function ${\cal
K}_{d,\sigma}(y)$ is easily deduced from equation (\ref{finalK}). It is
given by
\begin{widetext}
\begin{equation}
\label{asK} {\cal K}_{d,\sigma}(y)\approx\left\{
\begin{array}{ll}
\frac{\sigma}{\pi^{d/2}} \Gamma\left(\frac d2\right)\zeta(d)-
\sigma \frac{|D_{d-1,\sigma}|}{d-1}y^{(d-1)/\sigma},
&0<d-1<\sigma,\\[.5cm]
\frac{\sigma}{\pi^{d/2}}\Gamma\left(\frac d2\right)\zeta(d)
-2y\left[(4\pi)^{\sigma/2}\sigma\Gamma[\sigma/2]\right]^{-1}(1-\ln
y),
\hspace{4cm} &\sigma=d-1, \\[.5cm]
\frac{\sigma}{\pi^{d/2}}\Gamma\left(\frac
d2\right)\zeta(d)-y\frac{\pi^{(d-1)/2}}{(2\pi)^\sigma}
\frac{C_{d,\sigma}}{\Gamma[\sigma/2]}+
\sigma\frac{|D_{d-1,\sigma}|}{d-1}y^{(d-1)/\sigma}, &0<\sigma<d-1.
\end{array}
\right.
\end{equation}
\end{widetext}

For large $y$ the asymptotic behavior of the function
${\cal K}_{d,\sigma}(y)$ is obtained by substituting the
large $x$ behavior of the functions $E_{\alpha,\beta}(x)$ (given in Eq.
(\ref{largez})) in the definition (\ref{f}). After some calculations one
ends  up with
\begin{subequations}\label{largeyasy}
\begin{equation}
{\cal K}_{d,\sigma}(y)\simeq a_{d,\sigma} y^{-1},
\end{equation}
where
\begin{equation}
a_{d,\sigma}=\frac{2^{1+\sigma}}{\pi^{d/2}}
\frac{\Gamma\left[(d+\sigma)/2\right]}
{|\Gamma[-\sigma/2]|}\zeta(d+\sigma).
\end{equation}
\end{subequations}


\begin{thebibliography}{99}
\bibitem{note1} In that case  both finite-size and van der Waals
forces give raise to a contribution to the free energy of the
wetting layer that depends on its thickness $L$ as $L^{-2}$ for
$d=3$.
\end{thebibliography}
\end{document}